# Realization of Maxwell's Hypothesis

A heat-electric conversion in contradiction to Kelvin's statement

Xinyong Fu, Zitao Fu    Shanghai Jiao Tong University    Email: xyfu@sjtu.edu.cn

## ABSTRACT

Two similar and parallel Ag-O-Cs surfaces in a vacuum tube ceaselessly eject electrons at room temperature. A static magnetic field applied to the tube plays the role of "Maxwell's demon". The thermal electrons are so controlled by the magnetic field that they can travel only from one Ag-O-Cs surface to the other, resulting in collections of positive and negative charge on the two surfaces, respectively, as well as an electric potential between the two surfaces. A load, a resistance outside of the tube for example, is connected by wires to the two surfaces, continuously receiving electric power from the tube. The ambient air is a single heat reservoir in this situation, and all of the heat extracted by the tube from the air is converted to electric energy, without producing any other effect. The authors believe that the experiment is in contradiction to Kelvin's statement, and that the famous hypothesis proposed by Maxwell about 140 years ago is realized.

## 1. INTRODUCTION

In 1850 and 1851, Clausius and Kelvin established the famous second law of thermodynamics. According to this law, all practical thermodynamic processes known by mankind so far result in an increase of entropy, and they are all "irreversible". All energy conversions and energy transmissions that are related to thermal phenomena have a common single direction: a change from useful energy to useless one. Let us see an example: the electric energy that every family spends every day is, as we all know, useful energy, and every piece of such useful energy can be used only once. After being used, the electric energy has converted to waste heat and scatters into the surrounding air. Then it transmits to the farther atmosphere and ground, and eventually radiates to the cosmic space. The vast cosmic space is the terminal of all energy conversions and transmissions, and it is actually the final settlement of energy, the grave of energy.

The second law of thermodynamics, no matter whatever a statement it takes, always professes the same idea: energy cannot circulate, it just gets older and weaker without any backward step until its death, and after the death, it will never revive.

In 1871, to challenge this peculiar law, James Clerk Maxwell put forward an ingenious and creative hypothesis, the so-called (by Kelvin) Maxwell's demon [1]. The hypothesis goes directly into the possibility of energy refreshment, i.e., the possibility of energy circulation. It starts with a gas enclosed in an envelope which permits neither change of volume nor passage of heat, and in which both the temperature and pressure are everywhere uniform. There is a separator that divides the envelope into two equal parts, portion A and portion B, with a small door on the separator, as shown in Fig. 1. The demon's job is, without expenditure of work, to observe the motion of all the individual gas molecules and, according to the situation of the molecules' motion, open and close the door at proper times, so as to interfere intentionally with the molecules' random thermal motion. A demon may work in either of the two modes as below.

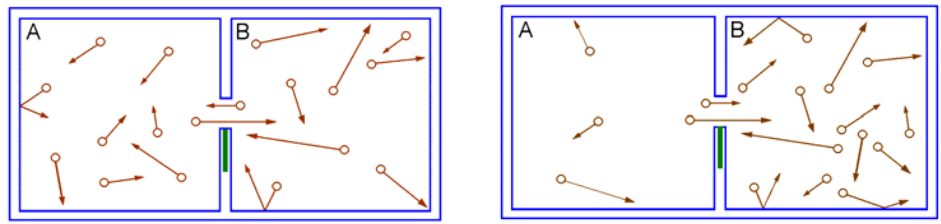

(a) In the first mode, the demon produces          (b) In the second mode, the demon
    inequality in temperature                          produces inequality in pressure

Fig. 1 How Maxwell's demon interferes with the random thermal motion of gas molecules*

In the first mode, as shown in Fig. 1(a), the demon allows only swifter molecules to pass the door from A to B, and slower ones from B to A, causing eventually a difference in temperature between A and B. (Note: this is a reverse process of heat transfer.)

In the second mode, as shown in Fig. 1(b), the demon allows the molecules, whether swifter or slower ones, to pass the door only from A to B, never from B to A, causing eventually a difference in pressure between A and B. (Note: this is a reverse process of gas free-expansion.)

Obtaining a difference in temperature or a difference in pressure without expenditure of work means the renewing of part of the energy of the gas [2], and that is in contradiction to the second law of thermodynamics.

*Note: Fig.1 of this paper is quoted from W. Ehrenberg's *Maxwell's Demon, Scientific American*, pp.103 (1967).

In the past 140 years, many people have endeavored to find a way to realize this attractive hypothesis. They tried with a variety of physical and other methods, nevertheless, all their efforts failed.

The authors hold that Maxwell's hypothesis will be much easier to realize if the demon, instead of dealing with the neutral molecules of a gas, turn to deal with the thermal electrons ejected by two cathodes in a vacuum tube [3].

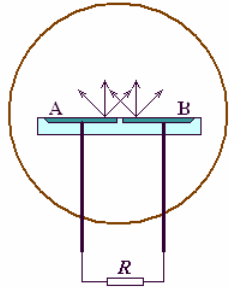

Fig. 2 Replacing Maxwell's neutral molecules with the thermal electrons
ejected by two cathode-material surfaces in a vacuum tube.

To illustrate this, let us imagine an electron tube whose essential part is an insulated plate (a quartz plate, for example) coated with two parallel and identical thermal electron ejectors on one of its surfaces, as shown in Fig. 2. We refer to the two ejectors after Maxwell as ejectors A and B. There is a narrow gap between A and B, keeping A and B close but well insulated from each other. The whole tube is immersed in some single-temperature heat reservoir whose temperature is such that A and B ceaselessly eject thermal electrons.

Fig. 3(a) illustrates the motion of electrons ejected from two points on A and B near the gap when there is no exterior magnetic field applied to the electron tube. Some of the electrons ejected by A can travel across the gap and fall on B, while an equal number of electrons ejected by B can also travel across the gap and fall on A. The two tendencies cancel each other, resulting in no net positive or negative charge on A or B.

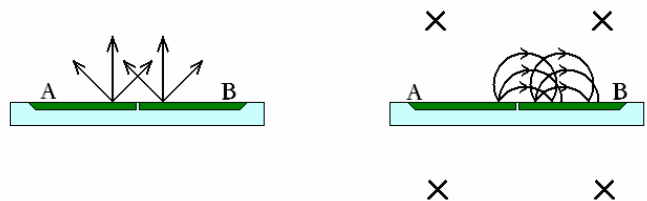

(a) Electrons ejected into a zero-field space    (b) Electrons ejected into a magnetic field
Fig. 3 How a magnetic demon deals with the thermal electrons.

Now, if we apply an appropriate static uniform magnetic field to the tube in the direction parallel to the gap between A and B, the paths of the ejected thermal electrons will change into circles with different radii, swifter electrons moving along larger circles and slower ones moving along smaller circles. As shown in Fig. 3(b), some of the electrons ejected by A can still travel easily across the gap and fall onto B, but it is now impossible for any electron ejected by B to travel across the gap and fall onto A. Such a net transition of electrons from A to B will rapidly result in a charge distribution, with A charged positively and B charged negatively. A potential difference between A and B is simultaneously created. The situation is very similar to the above mentioned second mode of Maxwell's demon, in which the gaseous molecules move only from portion A to portion B, resulting in a difference in pressure, as shown in Fig. 1(b).  Now, connect ejectors A and B with two metal wires to an external load, for example, a resistor or a reversible battery, the load will get a DC voltage and a DC current produced by the tube, both small but coherent and macroscopic.

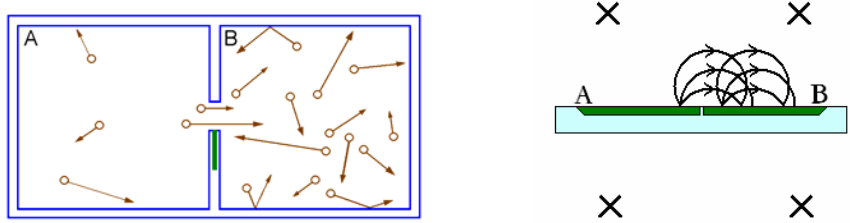

Fig.4 Compare the situations of Fig.1 (b) and Fig.3 (b) directly. The two cases are very similar from each other.

Such a voltage and current indicates a small electric power. The electron tube is supplying electric energy to the load continuously. One may ask, where this electric power comes from? It comes from the heat extracted by the electron tube from the heat reservoir in which the tube is immersed. We interpret the energy conversion from the heat extracted from the heat reservoir to electric energy as follows.

As mentioned above, when a magnetic field is applied to the tube, cathode A is rapidly charged positively while cathode B is charged negatively. These positive and negative charges produce a static electric field in the space above the gap between A and B. The direction of this electric field opposes the succeeding thermal electrons ejected by cathode A to travel towards cathode B. Nevertheless, the thermal electrons have kinetic energy. Relying on their kinetic energy, some of the thermal electrons ejected by A (especially the faster ones) will overcome the opposition of the static electric field and travel across the gap to fall onto B. On arriving at B, each electron has obtained a certain amount of electric potential energy, derived at the expense  of an equal amount of the electron's kinetic energy. Thus, the electrons are cooled down. Consequently the temperature of the electron tube will also fall down, slightly. The corresponding energy loss of the tube will be compensated automatically by the heat obtained from the surrounding heat reservoir.

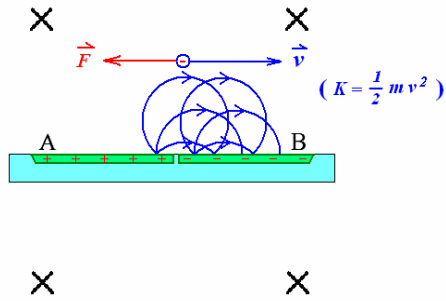

Fig.5  **F** is the static electric force exerted by the electric field between A and B, which is leftward, preventing the electrons to fly rightward; the speed of the electrons is rightward in general, the corresponding kinetic energy of the electrons help to overcome the prevention of **F** as they fly from A to B.

In the above process, the electron tube continuously extracts heat from a single-temperature heat reservoir and all of the extracted heat is converted into electric energy, without producing other effect. We maintain that the process is in contradiction to Kelvin's statement of the second law of thermodynamics.

The magnetic field here is actually a Maxwell's demon. It may be referred to as a magnetic demon.

By the way, besides Maxwell, Max Planck was also a brave challenger to the second law of thermodynamics. He claimed: the perpetual motion machine of the second kind was the best engine of mankind and would be mostly realized in some coming day. The above process confirms his genius prophecy [4].

The following is a detailed report of an actual experiment. We first describe the electron tube used in the experiment, focus on the cathode material and the tube structure, and then describe the performance of the experiment.

## 2. ELECTRON TUBE MODEL FX8

### A. The thermal electron ejectors and working temperature

Regarding the material of ejectors A and B, in principle, any of the known thermal electron cathode materials may be used for such an experiment. Nevertheless, most of the materials work at high temperature (≥800°C), if we were to adopt them, as the load and measuring instrument are usually at room temperature, there would be a large temperature difference along the closed electric circuit as shown in Fig. 2, causing serious disturbance to the test. To deal with these disturbances is difficult and complex. In order to accomplish a simple but integrated original experiment, we choose Ag-O-Cs cathode material. Ag-O-Cs has the lowest work function among all known thermal electron materials, only about 0.8 eV, and is currently optimum in maximizing thermal electron ejection at room temperature. We adopted this material and let the whole closed circuit (including the tube) work at room temperature, so that the whole closed circuit was readily kept at a uniform temperature, producing the least disturbance caused by thermo-electric effect. In a word, we wanted the test to be simple and ideal.

Ag-O-Cs cathodes are commonly used in photoelectric tubes and photo amplifiers. Due to their particularly low work function of 0.7 to 0.9 eV, Ag-O-Cs cathodes are highly sensitive to light, including near-infrared rays. As mentioned above, this low work function results in the greatest ejection of thermal electrons among all known cathodes materials at room temperature. The ejection of thermal electrons is usually referred to as dark current [5]. For common photoelectric tubes and photo amplifiers, users require the cathodes to produce minimal dark current. Most manufacturers adjust their technology to produce cathodes with very low dark current, usually in the range $10^{-11}$ to $10^{-14}$ A/cm$^2$. In our experiment, to the contrary, we require ejectors of maximum dark current. The authors, together with the engineers and workers of Yi Zhen Electron Tube Factory at Jiangsu Province, adjusted the manufacturing technology repeatedly over the past eleven years and succeeded in producing tubes with a three magnitudes stronger dark current, in the range $10^{-8}$ to $10^{-11}$ A cm$^{-2}$.

In this experiment, the tube plays the role of an electric power source. The load of the power source may be a resistor. In our present test, we used the input resistor of an electrometer, Keithley Model 6514, as the load. The electrometer was simultaneously used to measure the output current or voltage produced by the electron tube. The whole circuit, i.e. the power source (the tube) and the load (the input resistor of the electrometer), was kept at a uniform room temperature, so as to avoid Seebeck effect and other disturbances.

If instead of adopting Ag-O-Cs ejectors, we had adopted Ba-Sr-Ca oxide cathode material, the output current and power of the tube would certainly have increased dramatically (one million or ten million times greater). However, there would be prohibitive temperature differences in the closed circuit as the cathodes would be approximately at about 800°C while the load and electrometer would useually be at room temperature, say, 20°C. The great temperature difference along the closed circuit would result considerable Seebeck effect. In order to simplify the experiment, we abandoned Ba-Sr-Ca oxide and adopted Ag-O-Cs ejectors.

Another advantage of adopting Ag-O-Cs ejectors and room temperature is that the experiment extracts the "waste heat" from the ambient air in the laboratory and converts it directly to useful energy again. Such an experiment is certainly of great significance.

### B. The structure of the electron tube

The electron tube used in the experiment is FX8-24, whose structure is shown in Fig. 6. The outer envelope was of glass. A and B, the upper surfaces of two copper bases, were two similar and parallel Ag-O-Cs thermal electron ejectors, each with dimensions 4mm by 40 mm. Between the two copper bases there was a gap, that is, a mica sheet 0.07 ~ 0.09 mm in thickness, which kept A and B close together (about 0.10 mm apart) but well insulated from each other. Surfaces A and B and the top of the mica sheet were kept approximately in the same plane, as shown in Fig. 6 (a). At the bottom of A and B, the mica sheet was extended out about 5 ~ 7 mm, to prevent the electrons

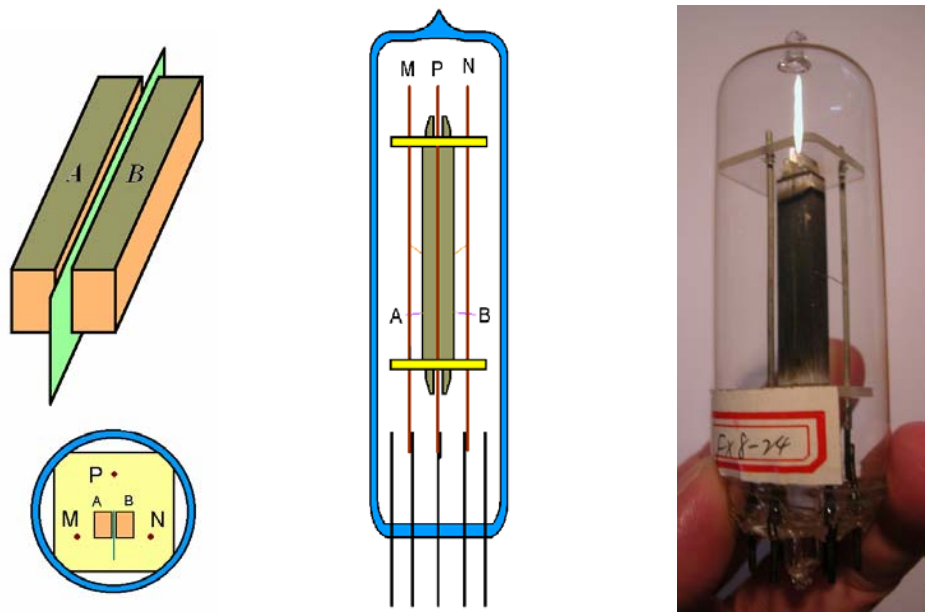

(a) Ejectors A and B, mica    (b) Sketch of the structure    (c) A photo of FX8-24
sheet, supports, rods P, M, N.      of tube FX8-24

Fig. 6  Electron tube FX8-24

from rotating back from B to A. M, N and P were three molybdenum supporting rods. M and N were also used as electrical leads connecting A and B to two of the tube base pins. P was 6mm above the gap separating A and B, and was used as a temporary anode in the process of manufacturing the Ag-O-Cs ejectors to enable oxidation of the silver films of A and B by oxygen-discharge. After completing of tube manufacture, P was also used temporarily as an anode to test whether the two ejectors were qualified for use in the destined experiment: A DC voltage of 30V was applied to the P-A circuit, and then separately to the P-B circuit, to measure the corresponding photoelectric sensitivity and dark current of ejectors A or B. The typical photoelectric sensitivity of each ejector was to be $0.2 \sim 10.0\mu A/lumn$, and the typical dark current of each ejector is $100 \sim 2500pA$. The leakage resistance between A and B was to be greater than $100\sim200M\Omega$. Outside of these ranges, according to our experience, a tube was unqualified for the destined experiment. The value of the leakage resistance depended chiefly on the input amount of cesium and the final heat activation and evacuation processes.

## 3. MEASUREMENT OF THE MAGNETIC FIELD AND THE OUTPUT CURRENT

### A. Measurement of the magnetic field

The magnetic field used to deflect the thermal electrons was produced by a 150 mm × 100 mm × 25 mm magnet (Ceramic 8, MMPA Standard). Fig. 7 shows the magnetic induction intensity $B$ at point O when the magnet is a distance $d$ from O. The $B \sim d$ relation was measured in advance with a tesla meter, see the results in Table I.

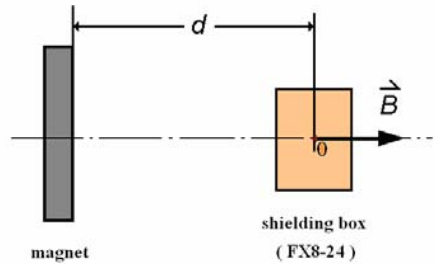

Fig. 7 The magnetic field produced by a single magnet used in the experiment

| $d_{cm}$ | 70 | 60 | 50 | 40 | 35 | 30 | 25 | 20 | 15 | 10 |
|---|---|---|---|---|---|---|---|---|---|---|
| $B\uparrow(N)_{gauss}$ | 0.2 | 0.3 | 1.1 | 2.1 | 2.9 | 4.4 | 7.2 | 13.1 | 25.5 | 59.7 |
| $B\downarrow(S)_{gauss}$ | -0.6 | -0.8 | - 0.7 | -1.6 | - 2.5 | - 4.0 | - 6.7 | - 12.7 | - 24.9 | - 58.8 |
| $B_{(mean), gauss}$ | 0.4 | 0.6 | 0.9 | 1.9 | 2.7 | 4.2 | 7.0 | 13 | 25 | 60 |

Table I $B \sim d$ relation of the magnet used in the experiment

## B. Output current and voltage

An electrometer of Keithley Model 6514 was used to measure the output current of electron tube FX8-24. The greatest current sensitivity of the electrometer was $1 \times 10^{-16}$A. A simple diagram of the measuring circuit is shown in Fig. 8.

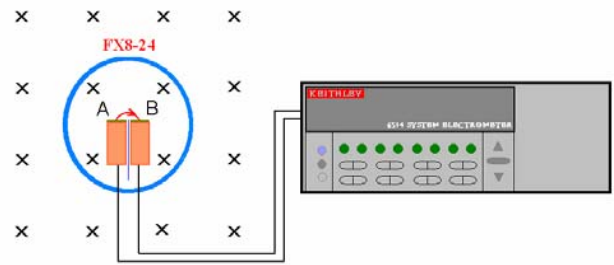

Fig.8 Current measuring circuit.

In addition to the output current, the instrument and circuit was also used to measure the output voltage.

## 4. THE EXPERIMENT AND RESULTS

Fig. 9(a) shows a photograph of the experiment set up. Prior to measurements, tube FX8-24 was placed inside a copper shielding box, as shown in Fig.9 (b). The four walls of the copper shielding box, together with its cover and base, were all 5 mm or more in thickness, so that external visible light and all other electromagnetic waves were excluded from the box interior. When a static magnetic field was applied t the tube, the anticipated output current or voltage of tube FX8-24 was transferred from the box

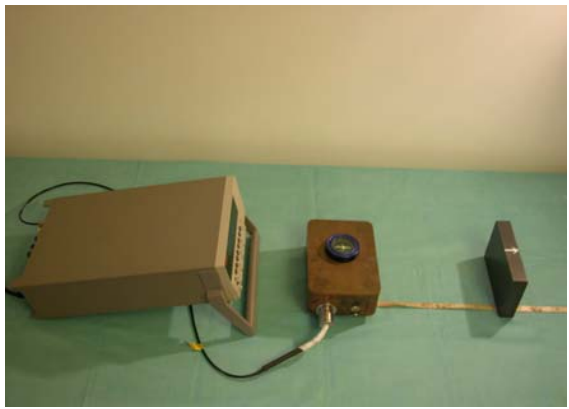
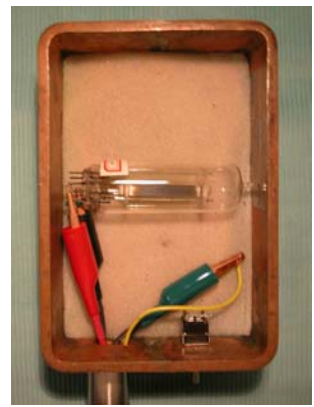

(a)

(b)

(a) Electrometer Keithley 6514, copper shielding box (containing FX8-24) and magnet.

(b) Position and orientation of electron tube FX8-24 in the copper shielding box.

Fig. 9 Set up of the experiment

to the electrometer through a concentric accessory cable, as shown in Fig.9 (a).

The output current produced by tube FX8-24 was measured at different magnetic induction intensities. As a start, we did not apply a magnetic field to the electron tube, $d = \infty, B \approx 0,$ and the electrometer showed satisfactory very weak readings typical of background noise. Then we applied to the tube a positive (northward) magnetic field $B_+$ of a relatively weak magnetic induction intensity, for example, $B_+ = 0.6$ gauss, corresponding to $d \approx 70$cm. A compass was placed on the top of the copper shielding box to show the direction of the magnetic field which should be aligned parallel to the axis of the tube (i.e., parallel to the gap between A and B). A positive output current emerged immediately. $B_+$ was then strengthened step by step by reducing $d$. A sufficiently long pause was given for each step to eliminate Faraday's motional electric motive force. A stable value of output current $I_+$ emerged for each step, corresponding to a magnetic induction intensity $B_+$. Initially, $I_+$ increased as $B_+$ was increased. There was then a peak in $I_+$, after which $I_+$ decreased as $B_+$ was increased further. Such a peak was according to our expectation and it is easy to be interpreted: Initially, as $B_+$ increased, the effect of the magnetic field that deflected the electrons to go only from A to B also strengthened, so the current increaseed in steps; after the magnetic field reached a certain value, the radii of the electron orbits became so small that a continually increasing number of electrons (the slower ones first) moved short to the left and were no longer capable of crossing the 0.10 mm gap between A and B, resulting in a continually decreasing current.

The magnet was then returned to $d > 70$cm and rotated 180° to reverse the direction of the magnetic field in the copper shielding box. The field then became negative (southward), and the corresponding magnetic induction intensity is now designated as $B_-$. As expected, the electrometer also showed a reversal in the direction of the output current. We designate the present current as $I_-$. As the distance $d$ was reduced in steps, $B_-$ increased correspondingly, and the reversed current $I_-$ first increased in magnitude, then decreased after a negative peak in a similar manner to that of $I_+$ after $B_+$.

We now refer to the currents $I_+$ and $I_-$ briefly as Maxwell's current, and denote it as $I$. For a given FX tube, Maxwell's current $I$ depends on two factors, the temperature $T$ and the magnetic induction intensity $B$,

$$I = I \, ( \, B \, , \, T \, ) \tag{1}$$

The detailed experimental results are shown in tabular and graphical form as follows.

Table II and Table III illustrate the data for the $I \sim B$ relation for electron tube FX8-24 measured in two tests. Fig.10(a) and (b) show the corresponding $I \sim B$ graphs. In each case, the temperature is regarded as a constant parameter.

| $d$ (cm) | ∞ | 50 | 40 | 35 | 30 | 25 | 20 | 15 |
|---|---|---|---|---|---|---|---|---|
| $B$ (gauss) | 0 | 0.9 | 1.9 | 2.7 | 4.2 | 7.0 | 13 | 25 |
| $I_+ (10^{-15} \text{A})$ | 0 | 7.4 | 11.4 | 13.4 | 15.1 | 11.3 | 3.0 | 2.1 |
| $I_- (10^{-15} \text{A})$ | 0 | -8.5 | -11.2 | -13.6 | -14.5 | -11.0 | -4.3 | -3.4 |

Table II $I \sim B$ relation of FX8-24 ( Test 1, T = 297.5K, $t$ = 24.4$^{\circ}$C )

| $d$ (cm) | ∞ | 50 | 40 | 35 | 30 | 25 | 20 | 15 |
|---|---|---|---|---|---|---|---|---|
| $B$ (gauss) | 0 | 0.9 | 1.9 | 2.7 | 4.2 | 7.0 | 13 | 25 |
| $I_+ (10^{-15} \text{A})$ | 0 | 8.0 | 10.5 | 12.8 | 14.3 | 11.1 | 5.3 | 4.6 |
| $I_- (10^{-15} \text{A})$ | 0 | -6.9 | -9.1 | -11.4 | -12.7 | -11.8 | -4.2 | -1.2 |

Table III $I \sim B$ relation of FX8-24 ( Test 2, T = 297.1K, $t$ = 24.0$^{\circ}$C )

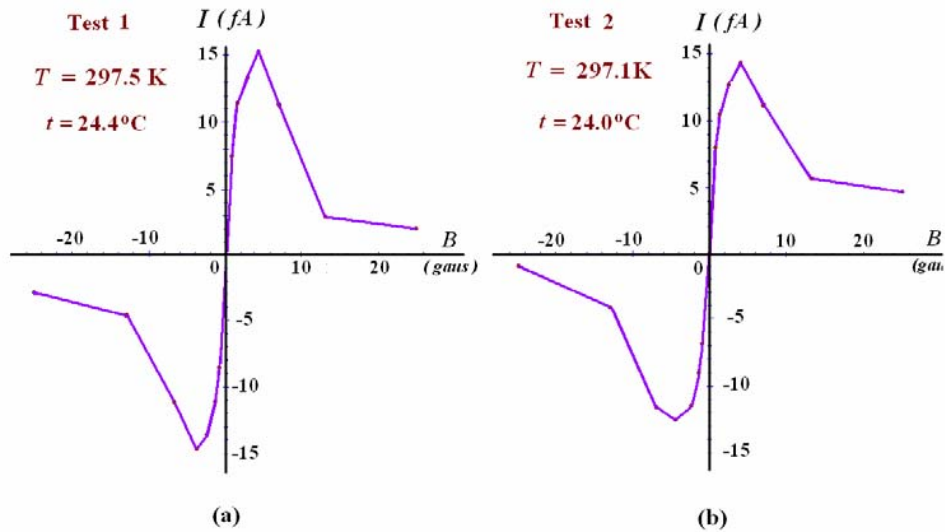

Fig.10 $I$ - $B$ graphs for electron tube FX8-24

The Keithley 6514 electrometer may also be used to measure the output voltage produced by tube FX8-24. Table IV and Table V list the data for voltage ~ magnetic induction intensity relation derived from two tests. The voltage here is the open-circuit voltage of FX8-24. Figure 11 (a) and (b) show the corresponding V ~ B graphs.

| $d$ (cm) | ∞ | 70 | 50 | 40 | 35 | 30 | 25 | 20 |
|---|---|---|---|---|---|---|---|---|
| $B$ (gauss) | 0 | 0.4 | 0.9 | 1.9 | 2.7 | 4.2 | 7.0 | 13 |
| $V_+$ (mV) | 0 | 0.07 | 0.22 | 0.30 | 0.29 | 0.25 | 0.10 | 0.03 |
| $V_-$ (mV) | 0 | -0.02 | -0.13 | -0.18 | -0.20 | -0.17 | -0.12 | -0.09 |

Table IV $V \sim B$ relation of electron tube FX8-24 (Test 1, T = 291.5K, $t$ = 18.4℃ )

| $d$ (cm) | ∞ | 70 | 50 | 40 | 35 | 30 | 27.5 | 25 |
|---|---|---|---|---|---|---|---|---|
| $B$ (gauss) | 0 | 0.4 | 0.9 | 1.9 | 2.7 | 4.2 | 5.2 | 7.0 |
| $V_+$ (mV) | 0 | 0.03 | 0.16 | 0.25 | 0.26 | 0.18 | 0.10 | 0.02 |
| $V_-$ (mV) | 0 | -0.07 | -0.14 | -0.20 | -0.21 | -0.16 | 0.12 | -0.07 |

Table V $V \sim B$ relation of electron tube FX8-24 (Test 2, T = 292.0K, $t$ = 18.9℃ )

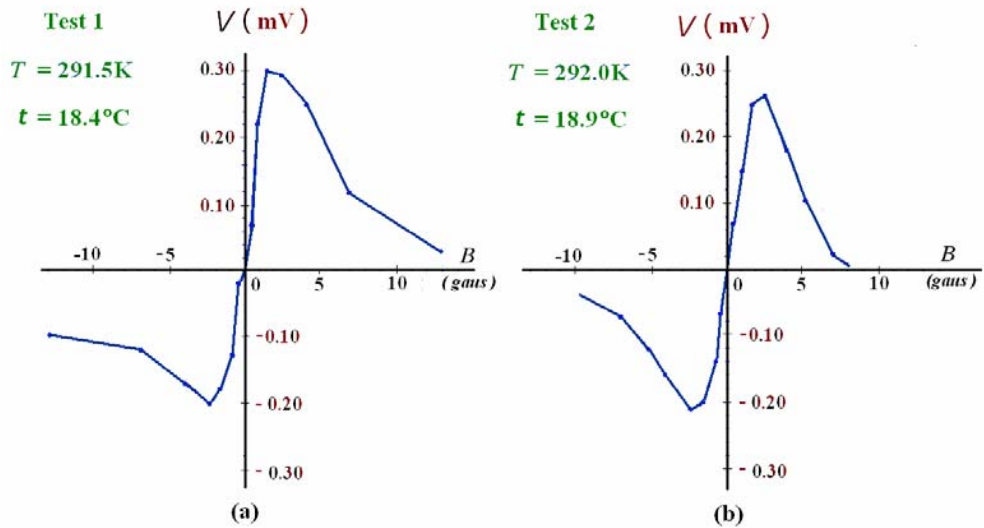

Fig.11 $V \sim B$ graphs for electron tube FX8-24.

The output currents and voltages of the experiment are very weak, but they are macroscopic currents and voltages. For example, the peak value of the output current in Fig. 10(a) is $I_+$ =15.1×10$^{-15}$A, the corresponding number of electrons passed from A to B per second is N = 15.1×10$^{-15}$A/1.6×10$^{-19}$ Col = 94000 electrons/sec. Such a unidirectional flow of electrons is undeniably a macroscopic direct current (DC). And, in Figure 11 (b), the positive and negative maximum values of the $V \sim B$ curve are $V_{+\ max}$ = 0.26 mV and $V_{-\ max}$ = 0.21 mV, both being explicitly macroscopic voltages.

Such macroscopic current and voltage are totally different from the thermal fluctuation current and voltage of the Johnson effect (i.e., thermal noise) in a metal bar.

A large number of such Maxwell current and voltage elements (Ag-O-C$_s$ pairs) may be connected in parallel to increase the current, and connected in series to increase the voltage, so as to form a much greater electric power. Because of this characteristic, we are more confident that Maxwell's current and voltage are macroscopic ones.

On the contrary, Johnson fluctuating currents and voltages are thermally random. Their directions alternate ceaselessly and irregularly, without any stable period or frequency. As is well known, they are impossible to be integrated to form a greater electric power. They are essentially thermal and random, i.e. microscopic.

## 5. CONCLUSIONS

In the above experiment, the heat transferred to the electron tube FX8-24 from the ambient air was converted completely into electric energy, without producing any other effect. The phenomenon proves clearly that the second law of thermodynamics is not universally valid.

The authors maintain: in ordinary thermodynamic processes, as pointed out correctly and profoundly by Clausius, entropy always increases, never deceases; nevertheless, in some extraordinary thermodynamic processes, such as our present experiment, entropy does decrease. These two kinds of thermodynamic processes are essentially different from each other, and they should be referred to as Clausius processes and Non-Clausius processes, respectively.

Thus, after a long, long delay, Maxwell's famous cherished wish has eventually been realized. Mankind has recognized a rather deeply hidden natural truth: Energy is actually capable of circulation.

Energy is immortal.

## ACKNOWLEDGEMENT

The authors are very grateful to Professors Fang Junxin and Hua Zhongyi, Professors Meng Zaoying and Qian Weichang; without their encouragement and help, we could not have progressed so well. Gu Weijia, an engineer in the Medical Equipment Center of Shanghai Jiao Tong University, and Xu Ping, an excellent engineer of Yi Zheng Electron Tube Factory, both helped us much in making the experimental electron tubes. We express here our sincere gratitude to them as well as to all the others who have helped us ardently.

APPENDIX

## 1. Maxwell's Original Hypothesis (Maxwell's Demon)

One of the best established facts in thermodynamics is that it is impossible in a system enclosed in an envelope which permits neither change of volume nor passage of heat, and in which both the temperature and the pressure are everywhere the same, to produce any inequality of temperature or of pressure without the expenditure of work. This is the second law of thermodynamics, and it is undoubtedly true as long as we can deal with bodies only in mass, and have no power of perceiving or handling the separate molecules of which they are made up. If we conceive a being whose faculties are so sharpened that he can follow every molecule in its course, such a being, whose attributes are still as essentially finite as our own, would be able to do what is at present impossible to us. For we have seen that the molecules in a vessel full of air at uniform temperature are moving with velocities by no means uniform, though the mean velocity of any great number of them arbitrarily selected, is almost exactly uniform. Now let us suppose that such a vessel is divided into two portions, A and B, by a division in which there is a small hole, and that a being, who can see the individual molecules, opens and closes this hole, so as to allow only the swifter ones to pass from A to B, and only the slower ones to pass from B to A. He will thus without expenditure of work, raise the temperature of B and lower that of A, in contradiction to the second law of thermodynamics.

**James Clerk Maxwell: *Theory of Heat* (**1871)

**2. Max Planck: the Legitimacy of the Perpetual Motion of the Second Kind**

*It is impossible to construct an engine which will work in a complete cycle, and produce no effect except the raising of a weight and the cooling of a heat reservoir.* Such an engine could be used simultaneously as a motor and a refrigerator without any waste of energy and material, and would in any case be the most profitable engine ever made. …… For this reason we take the above proposition as our starting point. Since we are to deduce the second law from it, we expect, at the same time, to make a most serviceable application of any natural phenomenon which may be discovered to deviate from the second law. As soon as a phenomenon is found to contradict any legitimate conclusion from the second law, this contradiction must arise from an inaccuracy in our first assumption, and the phenomenon could be used for the construction of the above-described engine. (§116)

In conclusion, we shall briefly discuss the question of the possible limitations to the second law. If there exist any such limitations - a view still held by many scientists and philosophers - this much may be asserted, that their existence presupposes an error in our starting point, viz. the impossibility of perpetual motion of the second kind, or a fault in our method of proof. From the beginning we have recognized the legitimacy of the first of these objections, and it cannot be removed by any line of argument. The second objection …… proves untenable. (§136)

In the mean time, no more effective weapon can be used by both champions and opponents of the second law than indefatigable endeavor to follow the real purport of this law to the utmost consequences, taking the latter one by one to the highest court of the appeal - experience. Whatever the decision may be, lasting gain will accrue to us from such a proceeding, since thereby we serve the chief end of natural science - the enlargement of our stock of knowledge. (§136)

**Max Planck: Treatise on Thermodynamics (**1897~ 1922)

**Video of The Experiment**

Please find the video of this experiment by the web link as following:
http://www.youtube.com/watch?v=FCCPeEKIVvQ&feature=youtu.be